\documentclass[structabstract]{aa}  
\usepackage{graphicx}
\usepackage{txfonts}
\usepackage{xcolor}
\usepackage[normalem]{ulem}

\begin{document}
   \title{Nonextensive distributions of rotation periods and diameters of asteroids}

   \titlerunning{Nonextensive distribution of asteroids}

   \author{Alberto S. Betzler
          \inst{1}
          \and
          Ernesto P. Borges\inst{2}
          }

   \authorrunning{A.\ S.\ Betzler \and E.\ P.\ Borges}

   \institute{Programa de Engenharia Industrial, Escola Polit\'ecnica \\
Universidade Federal da Bahia,
R. Aristides Novis 2, Federa\c{c}\~ao, 40210-630 Salvador--BA, Brazil \\
              \email{betzler.ssa@ftc.br}
         \and
Instituto de F\'{\i}sica 
 and
National Institute of Science and Technology for Complex Systems \\
Universidade Federal da Bahia,
Campus Universit\'ario de Ondina 40210-340 Salvador--BA, Brazil \\
             \email{ernesto@ufba.br}
             }

%   \date{}

% \abstract{}{}{}{}{} 
% 5 {} token are mandatory
 
  \abstract
   {To investigate the distribution of rotation periods of asteroids
    from  different regions of the Solar System
    and distribution of diameters of near-Earth asteroids (NEAs).}
   {Verify if nonextensive statistics satisfactorily describes the data.}
   {Light curve data was taken from Planetary Database System (PDS)
    with Rel $\ge 2$.
    Taxonomic class and region of the Solar System was also considered.
    Data of NEA were taken from Minor Planet Center.}
   {The rotation periods of asteroids follow a $q$-Gaussian
    with $q=2.6$ regardless of taxonomy, diameter or region of the Solar System
    of the object.
    The distribution of rotation periods is influenced by observational bias.
    The diameters of NEAs are described by a $q$-exponential with $q=1.3$.
    According to this distribution, there are expected to be
    $994 \pm 30$ NEAs with diameters greater than 1~km.
    }
   {}

   \maketitle
%

%_____________________________________________________________________________

\section{Introduction}

Asteroids and comets are primordial bodies of the Solar System (SS).
The study of the physical properties of these objects may lead to a better
understanding of the processes of formation of the SS, and, by inference,
of the hundreds of exo-Solar systems already known.
Distribution of rotation periods and diameters of asteroids are two
parameters that may give pieces of information concerning 
the evolution of the SS.
The first attempt to describe histograms of rotation periods of asteroids 
was made by Harris \& Burns (\cite{harris-burns}).
This work and also others that have followed
have shown that rotation periods
of big asteroids ($D>$30--40~km) follow a Maxwellian distribution.
Harris \& Pravec (\cite{pravec-harris-2000}) have analyzed a sample with
984 objects and have confirmed that the distribution of rotation periods of
asteroids with diameters $D\ge 40$ km is Maxwellian, with 99\% of confidence,
though this hypothesis can be rejected at 95\% of confidence.
They suggest that objects within this diameter range are primordial,
or originated from collisions of primordial bodies.
It is known that
for median sized ($10<D\le 40$~km ) and small ($D< 10$ km) asteroids,
the distribution of rotation periods is not Maxwellian.
The analysis of the data suggests the existence of a
spin-barrier for the asteroids with diameters between hundreds of meters
and 10 km and with more than 11 rotations per day (d$^{-1}$)
(period of about 2.2 h).
The absence of a substantial quantity of asteroids with period less than 2.2~h
may be due to the low degree of internal cohesion of these objects.
The majority of the sample may contain rubble pile asteroids 
(Davis et al.\ \cite{davis-1979}, Harris \cite{harris-1996}) 
that are composed by fragments of rocks kept together by self-gravitation.
For objects below 0.2~km it is observed rotation periods smaller than 
the spin-barrier, suggesting that these objects have a high
internal cohesion, implying that they may be monolithic bodies.
The difficulty in the modelling of rotation periods of asteroids as a whole
may be associated to the combined action of many mechanisms 
such as collisions (Paolicchi, Burns \& Weidenschilling \cite{paolicchi}), 
gravitational interactions with planets 
(Scheeres, Marzari \& Rossi \cite{scheeres-2004}),
angular momentum exchange in binary or multiple asteroid systems 
(Scheeres (\cite{scheeres-2002})),
or torques induced by solar radiation, known as YORP effect
(from Yarkovsky--O'Keefe--Radzievskii--Paddack) (Rubincam \cite{rubincam}).
Particularly, YORP effect is strongly dependent on the shape and size
of the object and its distance to the Sun.

Near-Earth Asteroids (NEAs) is a subgroup of SS asteroids whose heliocentric 
orbits lead them close to the Earth's orbit.
More than 7000 NEAs are known up to 2011. 
The study of these objects is relevant once it may bring information
regarding the birth and dynamic evolution of the SS.
Also a special interest in these objects is related to the possibility of 
collision with the Earth with obvious catastrophic consequences
(Alvarez et al.\ \cite{alvarez-1980}).
They also may be potential sources of raw material for future space projects.

The evaluation of the number of asteroids per year that may reach the Earth
as a function of their diameters is essential for the determination of the
potential risk of a collision.
One of the first attempts to estimate this flux was done by Shoemaker et al.\
(\cite{shoemaker-1979}).

The impact flux may be taken from the accumulated distribution
of diameters of the NEAs. 
This distribution is indirectly obtained by the current
asteroid surveys, according to the absolute magnitude $H$.
The distribution of absolute magnitude $H$ is described by 
Jedicke, Larsen \& Spahr (\cite{jedicke-2002}),
\begin{equation}
 \log N = \alpha H + \beta,
 \label{eq:distrib-h}
\end{equation}
where $N$ is the number of objects, $\alpha$ is the ``slope parameter''
and $\beta$ is a constant.
This relation asymptotically models the observed distribution of $H$.
Departure from this power law is probably associated with the observational
bias due to physical and dynamical properties of the asteroids
(orbital elements, size, albedo), 
and instrumental limitations (CCD, detection software, among others).
So Eq. (\ref{eq:distrib-h}) may be used to describe a given population
of asteroids if a correction of the bias is made in the raw data.

The diameters of asteroids $D$ may be given in function of their absolute 
magnitudes and their albedos $p_{_V}$ according to 
Bowell et al.\ (\cite{bowell})
\begin{equation}
 D = 1329 \frac{10^{-H/5}}{\sqrt{p_{_V}}}.
 \label{eq:d-h}
\end{equation}
The albedo is the rate of superficial reflection and its value is
essential for the estimation of the diameters of the asteroids.
The values of the albedos of asteroids varies according to the superficial 
mineralogical composition (taxonomic complex) and object's shape. 
Typical values range from $0.06\pm 0.02$ for low albedo objects of C taxonomic 
complex up to $0.46\pm0.06$ for high albedo objects of V type  
(Warner, Harris \& Pravec \cite{warner-harris-pravec-2009}).

Combination of equations (\ref{eq:distrib-h}) and (\ref{eq:d-h})
leads to a power law behavior: 
\begin{equation}
 N(>D)=kD^{-b}.
 \label{eq:distrib-d}
\end{equation}
The parameters are estimated by Stuart (\cite{stuart-2003}), 
$b=1.95$ ($\alpha=b/5$, admitting the same albedo for all the sample)
$k=1090$, and $D$ is given in km.
According to this expression and taking into account the uncertainties
of the measures, Stuart \& Binzel (\cite{stuart-binzel}) have estimated that
there may exist $1090\pm 180$ objects with diameters equal to 
or greater than 1~km ($H=17.8$).

%_____________________________________________________________________________

\section{Nonextensive statistics}

In order to model the accumulated distribution of periods and diameters of
asteroids, we have applied results from Tsallis nonextensive statistics.
This choice comes from observational evidences that astrophysical systems
are somehow related to nonextensive behavior.
It is known that system with long-range interactions (typically gravitational 
systems) are not properly described by Boltzmann-Gibbs statistical mechanics
(Landsberg \cite{landsberg}).
Along the last two decades it has been continuously developed 
the nonextensive statistical mechanics that is a generalization
of Boltzmann-Gibbs statistical mechanics. 
Tsallis proposed in 1988 (Tsallis \cite{tsallis:1988}) 
a generalization of the entropy,
\begin{equation}
 S_q = k \frac{1-\sum_i^W p_i^q }{q-1},
\end{equation}
where $p_i$ is the probability of the i-th microscopical state,
$W$ is the number of states, $k$ is a constant (Boltzmann's constant)
and $q$ is the entropic index. As $q\to 1$ $S_q$ is reduced to
Boltzmann-Gibbs entropy $S_1 = - k \sum_i^W p_i \ln p_i$.
It was soon realized that the nonextensive statistical mechanics
could be successfully applied to self-gravitating systems:
Plastino \& Plastino (\cite{plastinos}) found a possible solution
to the problem of the existence of a self-gravitating system with 
total mass, total energy and total entropy simultaneously finite,
within a nonextensive framework.
Many examples of nonextensivity in astrophysical systems may be found.
We list some instances.
Nonextensivity was observed in the analysis of magnetic field at distant 
heliosphere associated to the solar wind observed by Voyager 1 and 2 
(Burlaga \& Vi\~nas \cite{burlaga-2005},
 Burlaga \& Ness \cite{burlaga-2009}
 Burlaga \& Ness \cite{burlaga-2010}).
The distribution of stellar rotational velocities in the Pleiades
open cluster was found to be satisfactorily modelled by a 
$q$-Maxwellian distribution (Soares et al.\ \cite{soares-2006}).
The problem of Jeans gravitational instability was considered
according to nonextensive kinetic theory 
(Lima, Silva \& Santos \cite{jaslima}).
Nonextensive statistical mechanics was also used to describe galaxy clustering
processes (Wuensche et al.\ \cite{reinaldo-2004})
and temperature fluctuation of the cosmic background radiation 
(Bernui, Tsallis \& Villela \cite{cosmic-background-1},
Bernui, Tsallis \& Villela \cite{cosmic-background-2}).
Fluxes of cosmic rays can be accurately described by distributions that
emerge from nonextensive statistical mechanics
(Tsallis, Anjos \& Borges \cite{cosmicrays}).
A list of more instances of applications of nonextensive statistical mechanics 
in astrophysical systems may be found in Tsallis (\cite{tsallis-springer}).

It is important to mention that the index $q$ has a physical
interpretation --- it expresses the degree of nonextensivity ---
and for some systems it can be determined {\em a priori} 
(based on dynamical properties).
In fact a nonextensive system is characterized by a $q$-triplet
and not just by a single $q$ 
(additional information can be found in Tsallis (\cite{tsallis-springer})).
Such triplet was already obtained in an astrophysical system
(Burlaga \& Vinhas \cite{burlaga-2005},
 Burlaga \& Ness \cite{burlaga-2009}
 Burlaga \& Ness \cite{burlaga-2010}).

Maximization of $S_q$ under proper constraints leads to distributions that 
are generalizations of those that appear within Boltzmann-Gibbs context. 
For instance, if it is required that the (generalized) energy of the system 
is constant (Curado \& Tsallis \cite{curado-ct})
then the probability distribution that emerges is a $q$-exponential,
\begin{equation}
 p(x) \propto \exp_q (-\beta_q x),
\label{eq:qexp-distribution}
\end{equation}
$\beta_q$ is the Lagrange multiplier 
(not to confound with $\beta$ of Eq.~(\ref{eq:distrib-h})).
The $q$-exponential function is defined as (Tsallis \cite{tsallis-quimicanova})
\begin{equation}
 \exp_q x = [1 + (1-q) x]_+^{\frac{1}{1-q}}.
\label{eq:qexp-definition}
\end{equation}
The symbol $[a]_+$ means that $[a]=a$ if $a>0$ and $[a]=0$ if $a\le 0$.
The $q$-exponential is a generalization of the exponential function,
that is recovered if $q\to 1$.
If the constraint imposes that the (generalized) variance of the distribution 
is constant, then the distribution that maximizes $S_q$ is a $q$-Gaussian 
(Tsallis et al.\ \cite{tsallis-levy}, Prato \& Tsallis \cite{tsallis-prato}),
\begin{equation}
 p(x) \propto \exp_q (-\beta_q x^2).
\label{eq:qgauss}
\end{equation}
The $q$-Gaussian recovers the usual Gaussian at $q=1$, and particular
values of the entropic index $q$ turn $p(x)$ into various known distributions,
as Lorentz distribution, uniform distribution, Dirac's delta
(see Tsallis et al.\ \cite{tsallis-levy}, 
Prato \& Tsallis \cite{tsallis-prato}, for details).

The Lagrange parameter $\beta_q$ also has a precise physical meaning.
Within the statistical mechanics context, the Lagrange parameter $\beta_q$
in  Eq.~(\ref{eq:qexp-distribution}) is related to the inverse of the 
temperature ($\beta_1=1/(kT)$ if $q=1$),
and in Eq.~(\ref{eq:qgauss}) it is related to the inverse of the variance
($\beta_1=1/(2\sigma^2)$ in normal diffusion, $\sigma^2$ is the variance).
Generally speaking, the inverse of the Lagrange parameters are associated 
to the finiteness of the first or second moments of the distribution.

\begin{figure}
\begin{center}
 \includegraphics[width=0.35\textwidth,keepaspectratio,clip]{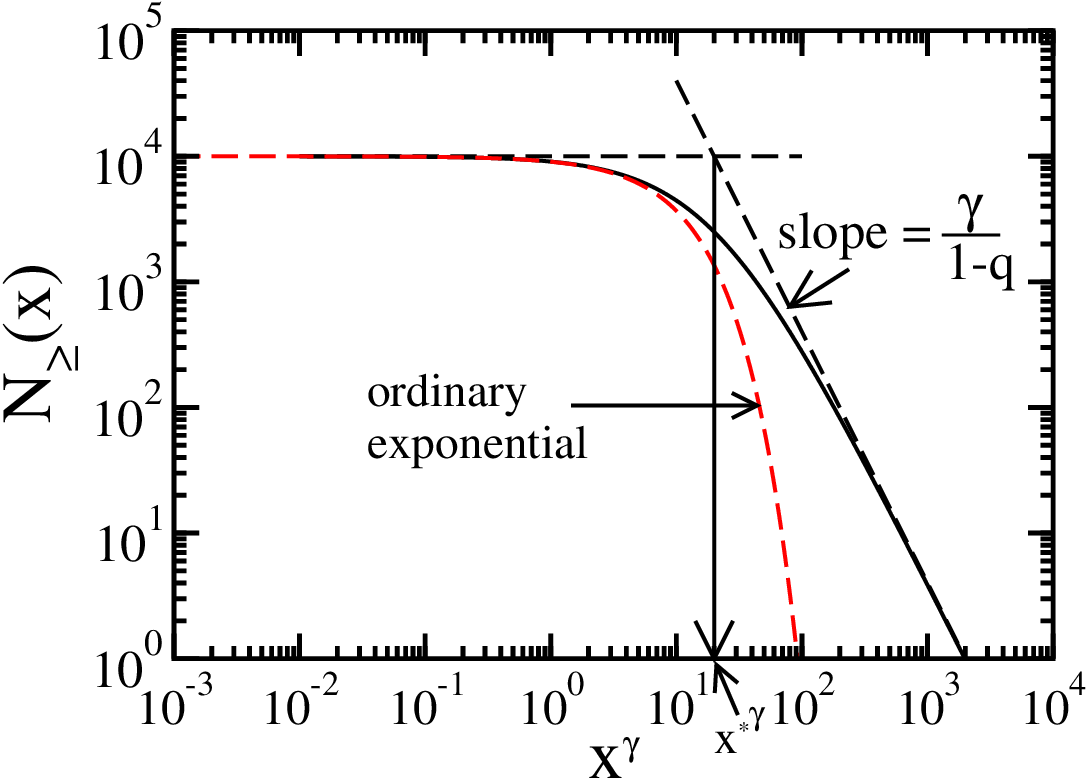}
 \caption{Inverse cumulative distribution of a generalized $q$-exponential,
          $N_{\ge}(x)= M \exp_q(-\beta_q x^\gamma)$. $\gamma=1$ 
	  is a $q$-exponential,
	  $\gamma=2$ is a $q$-Gaussian, $M$ is the size of the sample. 
          Values of the parameters for this particular instance are 
          $q=1.5$, $\beta_q=0.1$ and $M=10^4$.
          It is indicated the transition point $(x^*)^\gamma$ 
          given by Eq.\ (\protect\ref{eq:crossover}).
          Dashed curve (red on line) is an ordinary ($q=1$) exponential with
	  $\beta_1=0.1$, $\gamma=1$ and $M=10^4$. 
	  It is evident that the tail behavior is entirely different 
	  from both curves.
	  }
 \label{fig:q-distribution}
\end{center}
\end{figure}

Inverse cumulative distributions of the family of the $q$-exponentials 
are usually, and conveniently, graphically represented by means of 
$\log$-$\log$ plots, and Fig.\ \ref{fig:q-distribution} shows a general 
instance for the function $N_{\ge}(x)= M \exp_q(-\beta_q x^\gamma)$. 
$\gamma$ is a general parameter that recovers the $q$-exponential 
($\gamma=1$), the $q$-Gaussian ($\gamma=2$), and in a generalized scheme,
it may assume other values, i.e. it may represent other distributions.
The tails of $q$-exponentials are power laws
($\exp_q(-\beta_q x) \sim [(q-1)\beta_q x]^{1/(1-q)}$ for $x \gg 1$ and $q>1$)
and thus the slope of the asymptotical power law regime in a log-log plot
leads to the determination of the parameter $q$ ($\mbox{slope} = \gamma/(1-q)$ 
for general $\gamma$ not necessarily equal to one).
For small values of the independent variable $x$ (we are assuming $x>0$)
this graphical representation displays a region that appears to be 
quasi-flat in this $\log$-$\log$ plot; of course it is not flat, 
once the function is monotonically decreasing by construction. 
We are calling this region as ``quasi-flat'' to distinguish it 
from the asymptotical power law region.
Figure \ref{fig:q-distribution} shows the intersection of two straight lines
that represent the two regimes. The intersection is the transition point 
between the regimes (the crossover), and it is given by
\begin{equation}
x^*=\frac{1}{\left[(q-1)\beta_q\right]^{\frac{1}{\gamma}}}.
\label{eq:crossover}
\end{equation}
Figure \ref{fig:q-distribution} also displays an ordinary ($q=1$) exponential,
for comparison. Exponentials of negative arguments decay vary fast,
and when represented in $\log$-$\log$ plots this feature becomes clear,
once the exponential asymptotically presents  slope $\to-\infty$.
Coherently Eq.~(\ref{eq:crossover})  gives $x^*\to\infty$ for $q=1$, 
indicating that there is no crossover of an exponential to a power law regime.

We have found that the distribution of diameters of NEAs follows 
a $q$-exponential and the observed rotation periods of all asteroids,
regardless of their diameters, mineralogical composition or region of the SS,
are well approximated by a $q$-Gaussian.
We have used two samples from databases of different years, 
in order to verify the effect of the observational bias.

%_____________________________________________________________________________

\section{Observational data}

One important problem in the evaluation of distributions of
rotation periods and diameters of asteroids 
(and of course the same applies for other observables)
is that the data are possibly influenced by observational bias.
In order to take into account this effect, we have considered
samples from databases of two different years: 
2005 and 2010 for rotation periods,
and 2001 and 2010 for diameters of asteroids.

Two versions of the Lightcurve Derived Data, 
available at the Planetary Database System (PDS), were used:
version 7 (V7), with 1971 periods, and version 11 (V11), with 4310 periods
(Harris, Warner \& Pravec 
\cite{harris-warner-pravec-2005}, \cite{harris-warner-pravec-2010}).
The periods are classified according to a quality code
of the reliability of the estimated period, defined by 
Harris \& Young (\cite{harris-young}).
We have used periods with Rel $\ge 2$ (Rel from reliability)
that means they are accurate to $\approx 20\%$
which resulted in 
1621 entries for the V7 and 3567 asteroids for the V11.
Cross-checking the V11 sample with a compilation of taxonomic classifications,
also available at PDS, has revealed that about 40\% (1487)
of these asteroids have approximately known mineralogical composition.
The asteroids have been separated into three main classes: C , S and X 
complexes, following the SMASS II system of Bus \& Binzel (\cite{bus-binzel}),
with respectively 503, 663 and 321 objects.
The diameters of these sub-samples have been calculated with Eq.~(\ref{eq:d-h})
with the absolute magnitude $H$ available from MPCORB -- 
Minor Planet Center Orbit Database (MPC \cite{mpcorb}).

% NEAs
We have also used two versions of the compilation of absolute magnitudes $H$,
namely that of Oct., 2001, with 1649 NEA's
(similar to Stuart's (\cite{stuart-2001}) procedure)
and Oct., 2010, with 7078 objects.  (MPC \cite{mpcorb}). 
We have adopted $p_{_V}=0.14\pm 0.02$ for the NEAs population albedo.
This value was estimated by Stuart \& Binzel (\cite{stuart-binzel}) 
and it takes into account the great variety of taxonomic types that are found 
in the NEAs (Binzel et al.\ \cite{binzel-etal-2004}).
In order to estimate the validity of the diameters estimated by 
Eq.~(\ref{eq:d-h}), we have considered the diameters of 101 asteroids 
obtained from Spitzer Space Telescope data (Trilling et al.\ \cite{trilling}).
This resulted in about 20\% of error. We considered that this value, 
though not small, is reasonable for the purposes of our study.

%_____________________________________________________________________________

\section{Distribution of rotation periods}

\begin{figure}
\begin{center}
 \includegraphics[width=0.35\textwidth,keepaspectratio,clip]{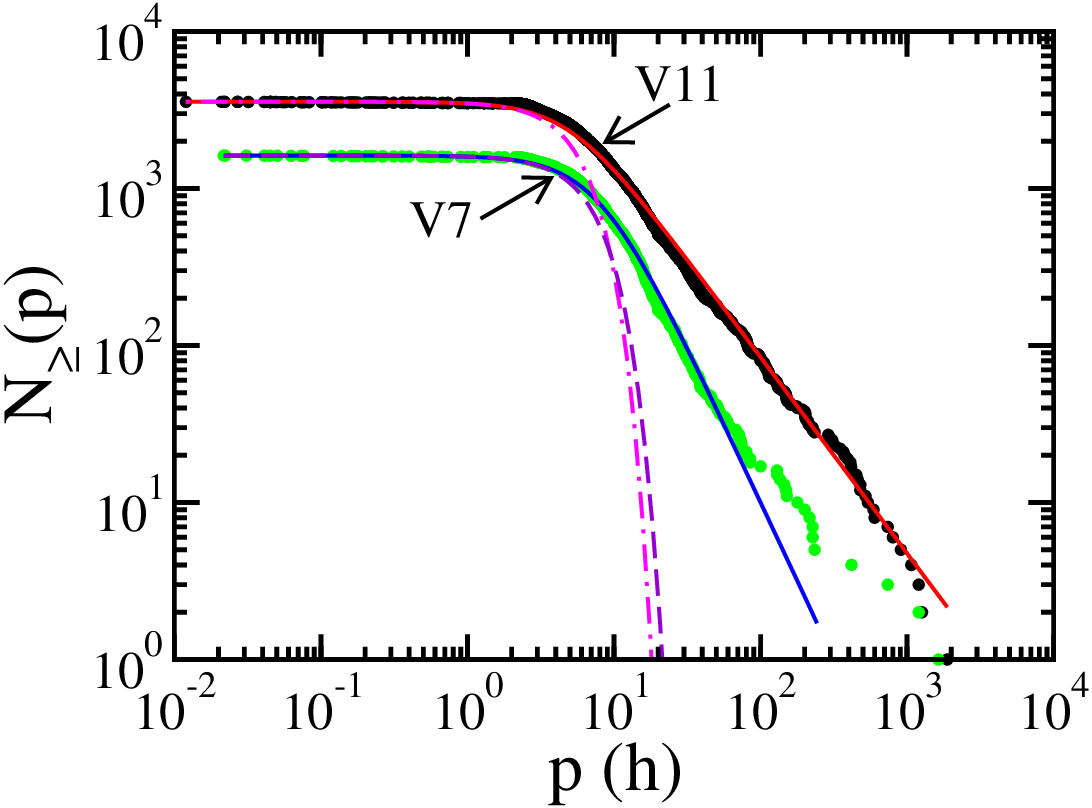}
 \caption{Decreasing cumulative distribution of periods of V7 
          (green dots on line) and V11 (black dots on line) of PDS (NASA) 
          with Rel $\ge 2$, and superposed $q$-Gaussians 
          ($N_{\ge}(p) = M \exp_q (-\beta_q p^2)$).
          V7:  $q=2.0$, $\beta_q = 0.0161$~h$^{-2}$, $M=1621$;
          V11: $q=2.6$, $\beta_q = 0.025$~h$^{-2}$, $M=3567$.
          Fittings of the periods of V7 for $P>50$~h are not good.
          This does not happen with V11, and it possibly indicates 
          the increase in the accuracy of the data.
          Dashed (violet on line) and dot-dashed (magenta on line) lines 
          are usual ($q=1$) Gaussians, with 
          $\beta_1 = 0.0161$~h$^{-2}$, $M=1621$, and
          $\beta_1 = 0.025$~h$^{-2}$, $M=3567$.
}
 \label{fig:periods-v7-v11}
\end{center}
\end{figure}

Fig.~\ref{fig:periods-v7-v11} shows the decreasing cumulative distribution of 
periods of V7 and V11, and  superposed $q$-Gaussians 
($N_{\ge}(p) = M \exp_q (-\beta_q p^2)$)
and it is seen that these functions quite satisfactorily describe almost all 
the data 
with $q=2.0\pm0.1$, $\beta_q=0.016\pm0.001$~h$^{-2}$ and $M=1621$
($M$ is the number of objects) for V7, 
and $q=2.6\pm 0.2$, $\beta_q = 0.025\pm 0.002$~h$^{-2}$ 
and $M=3567$ for V11.
Parameters were found by a nonlinear least square method.
Fig.~\ref{fig:periods-v7-v11} also presents two ordinary Gaussians,
and it is evident that these $q=1$ distributions are completely unable
to represent the data.
Confidence level for both fits is 95\%, according to $\chi^2$ test.
This suggests that the distribution does not depend on
(i) the diameters; 
(ii) the mineralogical composition; 
(iii) the region of the SS in which the object is found.
The latter is particularly important once the sample includes 
NEAs, trans-netunian objects (TNO), asteroids from the main belt (MBA), 
Jupiter Trojans (JT) and dwarf planets like Ceres and Pluto.
The values of the entropic indexes ($q=2.0$ for V7 and  $q=2.6$ for V11)
--- 
rather distant from unit, that is, distant from the Maxwellian distribution 
---
may indicate that long-range interactions play an essential 
role in the distribution of rotation periods.
According to Eq. (\ref{eq:crossover}) (with $x^* \equiv p^*$, $\gamma=2$),
the transition point is $p^*=7.91\pm0.01$~h
($f\sim 3$~d$^{-1}$) for the data of V7, 
and  $p^*=5.00\pm0.02$~h ($f\sim 5$~d$^{-1}$) for the data of V11.
The transition points for both samples differ from the critical period
of the spin barrier, and thus the transition is not a consequence
of physical processes.
Warner \& Harris (\cite{warner-harris-2010}) have demonstrated that the 
periods are more accurately determined for objects with periods $p\le8$~h
and light curve amplitudes $A\ge0.3$~mag, so we conclude that the difference 
between the transition points of the two versions is due to the action 
of observational bias.

\begin{figure}
\begin{center}
 \includegraphics[width=0.35\textwidth,keepaspectratio,clip]{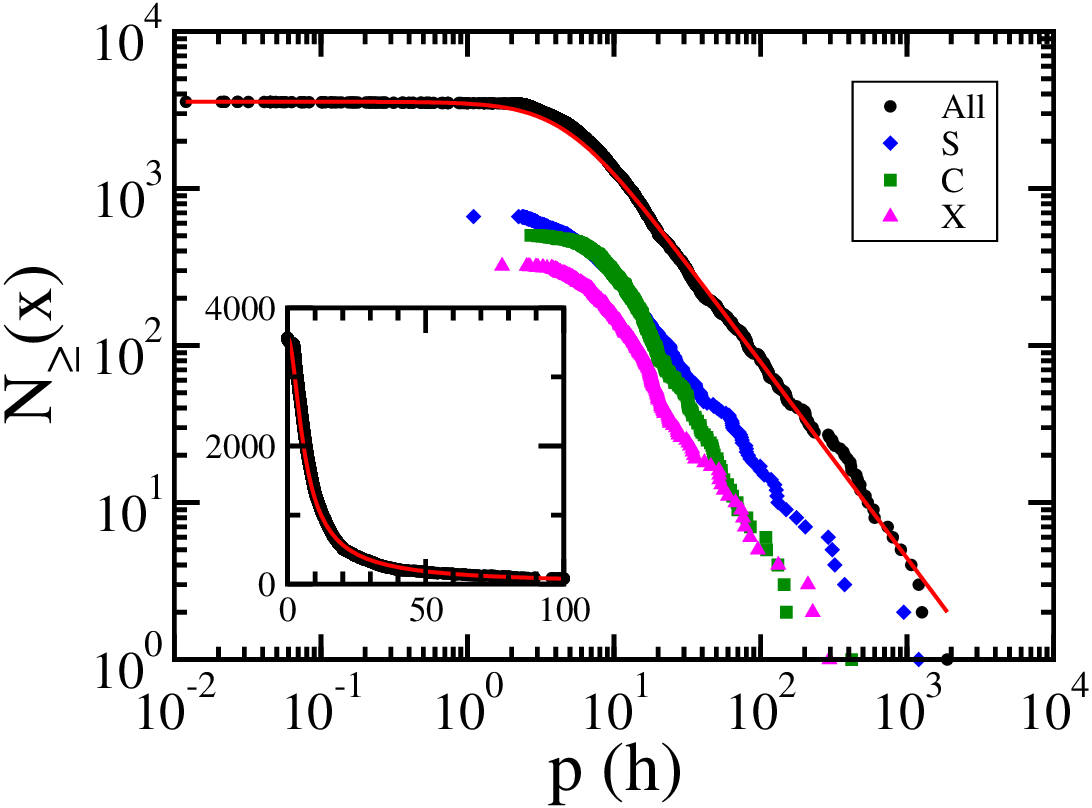}
 \caption{Log-log plot of the decreasing cumulative distribution of periods 
          of 3567 asteroids (dots) with Rel $\ge 2$
          taken from the PDS (NASA) and a $q$-Gaussian distribution
          ($N_{\ge}(p) = M \exp_q (-\beta_q p^2)$)
          (solid line), with $q=2.6$, $\beta_q=0.025$~h$^{-2}$, $M=3567$.
          The other curves are
          663 S-complex asteroids (diamonds, blue on-line), 
          503 C-complex asteroids (squares, green on-line), 
          321 X-complex asteroids (triangles, magenta on-line).
          Inset shows the 3567 asteroids and the $q$-Gaussian
          in a linear-linear plot.}
 \label{fig:periods-all-s-c-x}
\end{center}
\end{figure}

We have taken separately the taxonomic complexes C, S, and X (V11),
shown in Fig.~\ref{fig:periods-all-s-c-x}, 
and we have found that all of them are properly described by $q$-Gaussians
within 95\% of confidence level
($q = 2.6\pm0.2$ and $\beta_q = 0.021 \pm 0.002$~h$^{-2}$ for S, 
$q = 2.0\pm0.1$ and $\beta_q = 0.015 \pm 0.001$~h$^{-2}$ for C 
and $q = 2.0\pm0.1$ and $\beta_q = 0.010 \pm 0.007$~h$^{-2}$ for X;
fitted curves are not indicated in Fig.\ \ref{fig:periods-all-s-c-x}). 
The difference between the parameters for each complex is related to
the size of the sample, and they are statistically similar.
The fittings for the C and X complexes can be improved if we admit that
the number of objects are 10\% higher than that found in the sample.
The transition point $p^*$ gets smaller from V7 to V11, and this may be an 
indication that the fraction of fast rotators ($f\ge 5$~d$^{-1}$) may still 
be sub-estimated.

%_____________________________________________________________________________

\section{Distribution of diameters of near-Earth asteroids}

\begin{figure}
\begin{center}
 \includegraphics[width=0.35\textwidth,keepaspectratio,clip]{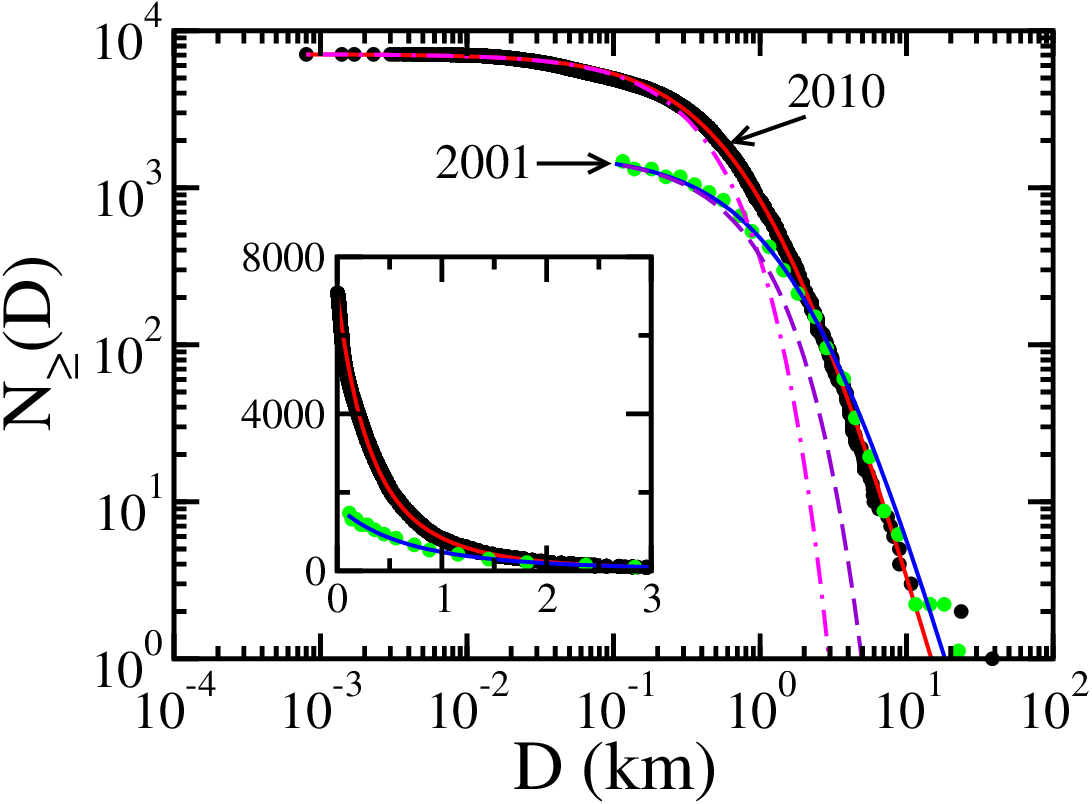}
\caption{Decreasing cumulative distribution of diameters of known NEAs in 2001 
         (1649 objects, green dots) and in 2010 (7078 objects, black dots).
         Solid lines are best fits of $q$-exponentials
         ($N_{\ge}(D) = M \exp_q (-\beta_q D)$).
         Blue line (2001): $q=1.3$, $\beta_q = 1.5$~km$^{-1}$, $M=1649$,
         red  line (2010): $q=1.3$, $\beta_q = 3$~km$^{-1}$, $M=7078$.
         Usual exponentials ($q=1$) are displayed in the main panel 
	 for comparison
         (dashed violet, with $\beta_1=1.5$~km$^{-1}$, $M=1649$,
         and dot-dashed magenta, with $\beta_1=3$~km$^{-1}$, $M=7078$).
	 }
\label{fig:diameters} 
\end{center}
\end{figure}

We have found that the decreasing cumulative distribution of diameters
of NEAs can be fitted by $q$-exponentials.
The fitting of a $q$-exponential to the diameters of 7078 NEAs
($N_{\ge}(D) = M \exp_q (-\beta_q D)$),
shown in Fig.~\ref{fig:diameters}, is quite good for the
entire range of the data, with a confidence level of 95\%: $q=1.3 \pm 0.1$ 
and $\beta_q = 3.0\pm 0.2$~km$^{-1}$ (found by nonlinear least squares method).
This distribution, however, is influenced by observational bias. 
The $q$-exponential distribution can be used to determine the point 
in which the sample is supposed to be complete.
Figure~\ref{fig:diameters} compares $q$-exponentials that fit observed
distribution of diameters of known NEAs in October of 2001 and October of 2010. 
The observed distribution of diameters of the 2011 database follows a
$q$-exponential with $q=1.3\pm 0.1$ and $\beta_q=1.5\pm 0.1$~km$^{-1}$
and the same confidence level.
The Figure also shows usual ($q=1$) exponentials, and it can be promptly 
verified their inadequacy in the representation of the whole range of the data.

Once the value of $q$ is the same for both 2001 and 2010 samples,
we may argue that this parameter is not influenced by the bias
in this case, and it reflects real physical processes.
Both curves are practically identical in the power-law region,
and the point of transition to the quasi-flat region differ, as expressed
by the different values of $\beta_q$.
The value of $q=1.3$ (different from one) is an indication that 
not only collisional processes are implied in the formation of these objects.
Other mechanisms may also be present: 
the YORP effect may lead to the decrease of the period of rotation 
up to the point of rupture.
This fragmentation process may yield the formation of binary 
or multiple systems.
About $(15 \pm 4)\%$ of NEAs with $D\ge 0.3$~km and rotation periods
between 2 and 3~h possibly are binary systems 
(Pravec, Harris \& Warner \cite{pravec-harris-warner-2007}).
The transition points according to Eq.\ (\ref{eq:crossover}) 
(with $x^* \equiv D^*$, $\gamma=1$)
are $D^*=2.22\pm0.05$~km (2001), and
$D^*=1.11\pm0.05$~km (2010). 

The sample is complete up to the upper limit of these intervals,
$2.22+0.05=2.27$~km ($H=16$) for 2001 basis and $1.11+0.05=1.16$~km ($H=17.5$)
for 2010 basis.
The number of NEA with $D\ge 2.27$~km is virtually the same for the sample of 
2001 and 2010 ($166\pm 8$ objects).
This is a confirmation of the completeness of the sample up to
$H\sim 15$ (Jedicke, Larsen \& Spahr, \cite{jedicke-2002}),
and the extension of this limit up to $H\sim 16$ (Harris, \cite{harris-2008}).
Once there has been an increase in the efficiency of detection and in the
number of surveys 
(Stokes, Evans \& Larson, \cite{stokes-2002}; Larson \cite{larson-2007}), 
we conclude that the parameter $\beta_q$ indicates the limit of completeness 
of the sample. For $D\ge 1.16$~km, the 2010 data are best described by 
a power-law. 
We have found for Eq.~(\ref{eq:distrib-d}),
$k=994\pm30$ and $b=2.24\pm0.01$, with correlation coefficient $R^2=0.987$.
The value of $b$ corresponds to $\alpha=0.448\pm0.002$ 
in Eq.~(\ref{eq:distrib-h}), that is a reasonable value if compared to
the slope of 0.44 found by Zavodny et al.\ (\cite{zavodny-2008}).
The value of $q$ may be found from $b$ using $q=1+1/b$, thus
$q=1.446\pm 0.001$, that is within the interval found in the whole sample.
According to the distribution we have found $994\pm30$ asteroids with
$D\ge1$~km ($H\le 17.7$), 
in a close agreement with Mainzer et al.\ (\cite{mainzer-2011}).
Analysis of distributions of diameters of MBA and TNO 
according to lines similar to those used in this work are welcome.

%_____________________________________________________________________________

\begin{acknowledgements}
This work was partially supported by FAPESB, through the program PRONEX
(Brazilian funding agency).
We are grateful to J.\ S.\ Stuart for important remarks and suggestions.
\end{acknowledgements}

%_____________________________________________________________________________

\end{document}